\documentclass[12pt]{article}

\def\eq#1{(\ref{#1})}

\begin{document}

\title{
Primordial Inflation and Present-Day Cosmological Constant
from Extra Dimensions}
\author{
Carl L. Gardner\thanks{Research
supported in part by the National
Science Foundation under grant DMS-9706792.}\\
{\em gardner@asu.edu}\\
Department of Mathematics\\ 
Arizona State University\\
Tempe AZ 85287\\ \ \\
{\em To appear in Physics Letters B}
}
\date{}

\maketitle
\thispagestyle{empty}

\begin{abstract}
A semiclassical gravitation model is outlined which makes use of the
Casimir energy density of vacuum fluctuations in extra compactified
dimensions to produce the present-day cosmological constant as
$\rho_\Lambda \sim M^8/M_P^4$, where $M_P$ is the Planck scale and $M$
is the weak interaction scale.  The model is based on
$(4+D)$-dimensional gravity, with $D = 2$ extra dimensions with radius
$b(t)$ curled up at the ADD length scale $b_0 = M_P/M^2 \sim$ 0.1 mm.
Vacuum fluctuations in the compactified space perturb $b_0$ very
slightly, generating a small present-day cosmological constant.

The radius of the compactified dimensions is predicted to be $b_0
\approx k^{1/4} 0.09$ mm (or equivalently $M \approx 2.4$
TeV/$k^{1/8}$), where the Casimir energy density is $k/b^4$.

Primordial inflation of our three-dimensional space occurs as in the
cosmology of the ADD model as the inflaton $b(t)$, which initially is
on the order of $1/M \sim 10^{-17}$ cm, rolls down its potential to
$b_0$.
\end{abstract}

\section{Introduction}

Supernova data indicate that the energy density $\rho_\Lambda$ in a
present-day cosmological constant is on the order of $0.7 \rho_c$,
where the current critical density $\rho_c \approx (2.5 \times 10^{-3}~
{\rm eV})^4$.  It is intriguing that $\rho_\Lambda \sim b_0^{-4}$
where $b_0 \sim$ 0.1 mm---just the length scale for compactified
extra dimensions predicted by Arkani-Hamed-Dimopoulos-Dvali (ADD) type
theories~\cite{ADD} with two extra spatial dimensions.

It is possible that this dark energy derives from vacuum fluctuations
in extra compactified dimensions.  We outline here a semiclassical
gravitation model which makes use of this mechanism to produce the
present-day cosmological constant.  The model is based on
$(4+D)$-dimensional gravity, with $D = 2$ extra dimensions with radius
$b(t)$ curled up at the ADD length scale $b_0$, where the subscript
``0'' denotes present-day values.

The ADD model can be realized~\cite{ADD-string} in type I
ten-dimensional string theory, with standard model fields naturally
restricted to a 3-brane~\cite{3-brane}, while gravitons propagate in
the full higher dimensional space.  For $D = 2$, two of the six
compactified dimensions are curled up with radius $\sim b_0$, while
the remaining four are curled up with radius $\sim 1/M_I$, with the
type I string scale $M_I \sim 1$ TeV\@.  In this picture, the ADD
model is formulated within a consistent quantum theory of gravity.  

In addition, if supersymmetry is broken only on the 3-brane, then the
bulk cosmological constant vanishes (see e.g.\
Ref.~\cite{exp-large-dim}).  A single fine tuning of parameters in the
potential for $b$ can then cancel the brane tension, setting the usual
four-dimensional cosmological constant to zero.

Semiclassical $(4+D)$-dimensional gravitation---with a potential for
the scale $b$ of the extra compactified dimensions---rapidly becomes a
good approximation to the string theory for energies below
$M_I$~\cite{hierarchy}.  In the semiclassical gravitation model, we
will assume a potential for $b(t)$ which stabilizes $b(t_0)$ at $b_0 =
M_P/M^2$ and which vanishes\footnote{In other words, we assume that
the 3-brane tension is exactly cancelled in the stabilization
potential at $b = b_0$.} at $b_0$ in the absence of the Casimir
effect, where the (reduced) Planck scale $M_P = 2.4 \times 10^{18}$
GeV and the weak interaction scale $M \sim 1$ TeV\@.  Vacuum
fluctuations in the compactified space will then perturb $b(t_0)$ very
slightly away from $b_0$, generating a small present-day cosmological
constant in our three-dimensional world.  This mechanism differs from
previous cosmological models incorporating the Casimir effect from
vacuum fluctuations in extra compactified dimensions (see e.g.\
Ref.~\cite{Candelas-Weinberg}), in which the Casimir energy density in
our three-dimensional world is cancelled by a bulk cosmological
constant.

Primordial inflation of our three-dimensional space will occur in the
model as the inflaton $b(t)$, which initially is on the order of $1/M
\sim 10^{-17}$ cm, rolls down its potential to
$b_0$~\cite{ADD-cosmo0,ADD-cosmo1}.  Many e-folds of inflation of our
3-space can occur for sufficiently flat potentials.

We will take the spacetime metric to be $R^1 \times S^3 \times T^2$
symmetric\footnote{Our treatment through Eq.~\eq{p} parallels that of
Kolb and Turner~\cite{Kolb-Turner}.}  where $ S^3$ is a 3-sphere and
$T^2$ is a 2-torus:
\begin{equation}
\label{metric}
	g_{M N} = {\rm diag}\{1, -a^2(t) \tilde{g}_{i j}, 
	-b^2(t) \tilde{g}_{m n}\}
\end{equation}
where $M,~ N$ run from 0 to 5; $i,~ j$ run from 1 to 3; and $m,~ n$
run from 4 to 5.  $\tilde{g}_{i j}$ is the metric of a unit 3-sphere
and $\tilde{g}_{m n}$ is the metric of a unit 2-torus, with $a(t)$ the
radius of physical 3-space and $b(t)$ the radius of the compactified
space.

The nonzero components of the $(4+D)$-dimensional Ricci tensor
are
\begin{eqnarray}
\label{Ricci}
	R_{00} & = & - \left( 3 \frac{\ddot{a}}{a} + D \frac{\ddot{b}}{b}
	\right)
\nonumber \\
	R_{ij} & = & - \left( \frac{\ddot{a}}{a} + 2 \frac{\dot{a}^2}{a^2}
	+ D \frac{\dot{a}}{a} \frac{\dot{b}}{b} 
	+ \frac{2}{a^2}	
	\right) g_{ij}
\nonumber \\
	R_{mn} & = & - \left( \frac{\ddot{b}}{b} 
	+ (D - 1) \frac{\dot{b}^2}{b^2}
	+ 3 \frac{\dot{a}}{a} \frac{\dot{b}}{b} 
	\right) g_{mn} .
\end{eqnarray}
The generalized Einstein equations are
\begin{equation}
\label{Einstein}
	R_{MN} = 8 \pi \overline{G} \left( T_{MN} - \frac{{T^P}_P}{D+2} 
	g_{MN} \right) 
\end{equation}
where $8 \pi \overline{G} = 8 \pi G \mathcal{V}_0 =
\mathcal{V}_0/M_P^2 = \tilde{\Omega}_D/M^{D+2}$ is the
$(4+D)$-dimensional gravitational constant, $\mathcal{V}_0 =
\tilde{\Omega}_2 b_0^2$ is the volume of the compactified dimensions
today, $\tilde{\Omega}_D$ denotes the volume of the unit $D$-torus,
and $T_{MN}$ is the energy-momentum tensor.  The gravitational
coupling $8 \pi G = 1/(b_0^2 M^4)$ is weak in the ADD picture because
$b_0$ is much greater than the $(4+D)$-dimensional Planck length
$1/M$.

The nonzero components of the energy-momentum tensor are given by
\begin{eqnarray}
\label{energy-momentum}
	T_{00} & = & \rho \nonumber \\
	T_{ij} & = & - p_a g_{ij} \nonumber \\
	T_{mn} & = & - p_b g_{mn} .
\end{eqnarray}
Thus ${T^P}_P = \rho - 3 p_a - D p_b$.  Expressed in terms
of the radii $a$ and $b$, the energy density $\rho$, and the
pressures $p_a$ and $p_b$, the Einstein equations become
\begin{equation}
\label{I-0}
	3 \frac{\ddot{a}}{a} + D \frac{\ddot{b}}{b} =
	- \frac{8 \pi \overline{G}}{D+2} \left[
	(D+1) \rho + 3 p_a + D p_b
	\right]
\end{equation}
\begin{equation}
\label{I-1}
	\frac{\ddot{a}}{a} + 2 \frac{\dot{a}^2}{a^2}
	+ D \frac{\dot{a}}{a} \frac{\dot{b}}{b} 
	+ \frac{2}{a^2} =
	\frac{8 \pi \overline{G}}{D+2} \left[
	\rho + (D-1) p_a - D p_b
	\right]
\end{equation}
\begin{equation}
\label{I-2}
	\frac{\ddot{b}}{b} + (D - 1) \frac{\dot{b}^2}{b^2}
	+ 3 \frac{\dot{a}}{a} \frac{\dot{b}}{b} 
	= \frac{8 \pi \overline{G}}{D+2} \left[
	\rho - 3 p_a + 2 p_b
	\right] .
\end{equation}
After a few e-folds of primordial inflation of our physical 3-space,
the curvature term $2/a^2$ on the left-hand side of Eq.~\eq{I-1} will
be negligible, and we will henceforth set this term to zero.  

We will be looking for solutions (neglecting matter) in which physical
3-space is inflating at the present epoch during which $b(t)$ is fixed
at $b_0$, or in the primordial epoch just after the quantum birth of
the universe during which $b(t)$ is inflating to its present value.
For an inflating 3-space (without matter), $p_a = - \rho$ and the
Einstein equations become
\begin{equation}
\label{Einstein-0}
	3 \frac{\ddot{a}}{a} + D \frac{\ddot{b}}{b} =
	\frac{8 \pi \overline{G}}{D+2} \left[
	-(D-2) \rho - D p_b 
	\right]
\end{equation}
\begin{equation}
\label{Einstein-1}
	\frac{\ddot{a}}{a} + 2 \frac{\dot{a}^2}{a^2}
	+ D \frac{\dot{a}}{a} \frac{\dot{b}}{b} =
	\frac{8 \pi \overline{G}}{D+2} \left[
	-(D-2) \rho - D p_b 
	\right]
\end{equation}
\begin{equation}
\label{Einstein-2}
	\frac{\ddot{b}}{b} + (D - 1) \frac{\dot{b}^2}{b^2}
	+ 3 \frac{\dot{a}}{a} \frac{\dot{b}}{b} 
	= \frac{8 \pi \overline{G}}{D+2} \left[
	4 \rho + 2 p_b
	\right] .
\end{equation}

The energy density and pressures on the right-hand sides of
Eqs.~\eq{Einstein-0}--\eq{Einstein-2} are derivable from the internal
energy $U = U(a, b)$:
\begin{equation}
\label{p}
	\rho = \frac{U}{\mathcal{V}} ,~~
	p_a = - \frac{a \partial U/\partial a}{3 \mathcal{V}} ,~~
	p_b = - \frac{b \partial U/\partial b}{D \mathcal{V}}
\end{equation}
where $\mathcal{V} = \Omega_3 a^3 \tilde{\Omega}_2 b^2$ is the volume
of $(3+D)$-space and $\Omega_3$ denotes the volume of the unit
3-sphere.

We will consider a potential $V(b)$ for the radius $b(t)$ in the
internal energy
\begin{equation}
\label{U}
	U(a, b) = \Omega_3 a^3 M^4 V(b)
\end{equation}
(at zero temperature) which will produce sufficient primordial
inflation to solve the horizon, flatness, homogeneity, isotropy, and
monopole problems, and which will stabilize $b$ at $b_0 = M_P/M^2
\sim$ 0.1 mm, with a vanishing cosmological constant.  Note that if
$p_a$ is to equal $-\rho$, then $U$ must be proportional to $a^3$, and
that $V(b)$ is dimensionless.

The potential $V(b)$ will generate a potential $B(b)$ with the
right-hand side of the Einstein equation~\eq{Einstein-2} equal to
$-B'(b)/b$.  If $B(b)$ is sufficiently flat near $b \sim 1/M$, then
many e-folds of inflation will occur in our physical 3-space as $b(t)$
rolls from $1/M$ to $b_0$.

Quantum fields will be periodic in the compactified space, producing a
Casimir effect~\cite{Candelas-Weinberg} in the compactified space and
in our three-dimensional world.  Adding a Casimir (C) term to the
internal energy
\begin{equation}
\label{UC}
	U_C(a, b) = \Omega_3 a^3 \left( \frac{k}{b^4} + M^4 V(b) \right)
\end{equation}
from vacuum fluctuations in the compactified space will perturb
$b(t_0)$ very slightly away from $b_0$ and generate a residual
present-day cosmological constant $\rho_\Lambda = k/b_0^4$.  The sign
and magnitude\footnote{A logarithmic dependence $\ln(M^2 b^2)$ can be
absorbed into the definition of $k$ without changing the conclusions
below.} of the constant $k$ depend on the particle content and
structure of the underlying quantum gravity theory.  The magnitude of
$k$ may be expected to be roughly in the range $10^{-7}$--$10^{-3}$
based on the analysis of Candelas and
Weinberg~\cite{Candelas-Weinberg}, who calculated the one-loop Casimir
contribution from massless scalar and spin-$\frac{1}{2}$ particles in
$(4+D)$-dimensional gravitation with an odd number of extra dimensions
$D$ curled up near the Planck length.  In their work, $k$ is positive
for a single massless real scalar field for odd dimensions $3 \le D
\le 19$, but may be positive or negative.  For our model to produce a
positive present-day cosmological constant, we will need $k > 0$.

\section{Primordial Inflation}

In this section, we briefly review the cosmological results for
primordial inflation of Refs.~\cite{ADD-cosmo0,ADD-cosmo1} for the ADD
model with internal energy $U$, and check that the Casimir terms in
the Einstein equations when $U$ is replaced by $U_C$ do not
qualitatively change the primordial cosmological picture.

The Einstein equations with the internal energy given by $U$ in
Eq.~\eq{U} take the form
\begin{equation}
\label{II-0}
	3 \frac{\ddot{a}}{a} + 2 \frac{\ddot{b}}{b} = 
	3 \dot{H} + 3 H^2 + 2 \dot{H}_b + 2 H_b^2 =
	\frac{V'(b)}{4 b}
\end{equation}
\begin{equation}
\label{II-1}
	\frac{\ddot{a}}{a} + 2 \frac{\dot{a}^2}{a^2}
	+ 2 \frac{\dot{a}}{a} \frac{\dot{b}}{b} =
	\dot{H} + 3 H^2 + 2 H H_b =
	\frac{V'(b)}{4 b}
\end{equation}
\begin{equation}
\label{II-2}
	\frac{\ddot{b}}{b} + \frac{\dot{b}^2}{b^2}
	+ 3 \frac{\dot{a}}{a} \frac{\dot{b}}{b} =
	\dot{H}_b + 2 H_b^2 + 3 H H_b =
	\frac{V(b)}{b^2} - \frac{V'(b)}{4 b} \equiv - \frac{B'(b)}{b}
\end{equation}
where the Hubble parameters $H \equiv \dot{a}/a$ and $H_b \equiv
\dot{b}/b$.  For a vanishing present-day cosmological constant,
$V'(b_0) = 0$ from Eq.~\eq{II-1}.  Eq.~\eq{II-2} then implies $V(b_0)
= 0$ to stabilize $b(t_0)$ at $b_0$.  

To summarize the successful phenomenology of Ref.~\cite{ADD-cosmo1}:
The ADD model can produce sufficient inflation ($\gg 70$ e-folds) to
solve the cosmological problems for a class of potentials $V(b)$ which
satisfy
\begin{equation}
	H^{-1} \sim H_b^{-1} \ge \frac{1}{M}
\end{equation}
at the beginning of inflation at the quantum birth of the universe
when $a \sim b \sim 1/M$, and
\begin{equation}
	H \gg H_b ,~~ \dot{H}_b \ll H^2
\end{equation}
during the initial stages of inflation.  The correct magnitude and
approximate scale invariance of density perturbations $\delta
\rho/\rho = 2 \times 10^{-5}$ are created if at an intermediate stage
of inflation when $b(t) \sim 10^{3/2}/M \ll b_0$, $H_b \approx H/100$.
There may be a period of contraction (similar to the vacuum Kasner
solutions) of our physical 3-space, but for $D = 2$, the amount of
contraction of $a(t)$ is at most 7 e-folds, so the contraction phase
does not invalidate the solution of the flatness problem.

Replacing $U$ by $U_C$ in Eq.~\eq{UC} introduces Casimir terms into
the Einstein equations:
\begin{equation}
\label{IIk-0}
	3 \dot{H} + 3 H^2 + 2 \dot{H}_b + 2 H_b^2 =
	-\frac{k}{M^4 b^6} + \frac{V'(b)}{4 b}
\end{equation}
\begin{equation}
\label{IIk-1}
	\dot{H} + 3 H^2 + 2 H H_b =
	-\frac{k}{M^4 b^6} +
	\frac{V'(b)}{4 b}
\end{equation}
\begin{equation}
\label{IIk-2}
	\dot{H}_b + 2 H_b^2 + 3 H H_b =
	\frac{2 k}{M^4 b^6} +
	\frac{V(b)}{b^2} - \frac{V'(b)}{4 b} \equiv - \frac{{B_C}'(b)}{b} .
\end{equation}
The Casimir terms do not qualitatively change the primordial
inflationary period of the ADD model, since initially
\begin{equation}
	\frac{k}{M^4 b^6} \approx k M^2 \ll M^2 \sim H^2 \sim H_b^2
\end{equation}
and in the intermediate stage of inflation
\begin{equation}
	\frac{k}{M^4 b^6} \approx 10^{-9} k M^2 \ll 10^{-11} M^2 \sim
	H_b^2 \sim 10^{-4} H^2 
\end{equation}
for $k {\stackrel{<}{\mbox{\scriptsize $\sim$}}} 10^{-3}$, using the
estimates in Ref.~\cite{ADD-cosmo1}.

\section{Present-Day Cosmological Constant}

In the present epoch, the internal dimensions have a fixed radius
$b(t_0) \gg 1/M$ and $H_b = 0$.  Without the Casimir terms, the static
solution for $b(t_0)$ requires $V(b_0) = 0 = V'(b_0)$.  In our model,
vacuum fluctuations in the compactified space perturb $b_0$ very
slightly to $\tilde{b}_0$, producing a small cosmological constant in
our three-dimensional world.  We assume that the potential $V(b)$ is
independent of the Casimir effect, so that $V(b_0)$ and $V'(b_0)$
still equal zero.

The Einstein equations with Casimir contributions for an inflating
3-space now take the form
\begin{equation}
\label{a}
	3 H_0^2 =
	-\frac{k}{M^4 \tilde{b}_0^6} +
	\frac{V'\left(\tilde{b}_0\right)}{4 \tilde{b}_0}
\end{equation}
\begin{equation}
\label{b}
	0 = \frac{2 k}{M^4 \tilde{b}_0^6} +
	\frac{V\left(\tilde{b}_0\right)}{\tilde{b}_0^2} -
	\frac{V'\left(\tilde{b}_0\right)}{4 \tilde{b}_0} .
\end{equation}
Setting $\tilde{b}_0 = (1 + \delta) b_0$ and solving Eq.~\eq{b} to
order $\delta \sim M^4/M_P^4$ yields
\begin{equation}
	\frac{\delta}{4} V''(b_0) + O(\delta^2) = \frac{2 k}{M^4 b_0^6}
\end{equation}
or
\begin{equation}
	\tilde{b}_0 \approx 
	\left( 1 + \frac{8 k}{M^4 b_0^6 V''(b_0)} \right) b_0
	= \left( 1 + O\left( \frac{k M^4}{M_P^4} \right) \right) b_0
\end{equation}
where $V''(b_0) \sim 1/b_0^2 = M^4/M_P^2$.  Eq.~\eq{a} then predicts a
present-day cosmological term
\begin{equation}
	3 H_0^2 = \frac{\delta}{4} V''(b_0) - \frac{k}{M^4 b_0^6} + O(\delta^2)
	= \frac{k}{M^4 b_0^6} + O(\delta^2)
\end{equation}
 or, in other words, 
\begin{equation}
\label{0}
	H_0^2 =
	\frac{8 \pi G}{3} \rho_\Lambda ,~~
	\rho_\Lambda = \frac{k}{b_0^4} = \frac{k M^8}{M_P^4} .
\end{equation}
This cosmological term will $\approx 0.7 \rho_c$ if $b_0 \approx
k^{1/4} 0.09$ mm, or equivalently if $M \approx 2.4$ TeV/$k^{1/8}$.

Note that the Casimir effect has caused the stabilized radius $b_0$ to
{\em increase} slightly, yielding a positive present-day cosmological
constant.

The canonically normalized ``radion'' field $\varphi(t) = 2 M^2 b(t)$.
The mass squared of the radion field is
\begin{equation}
	m_\varphi^2 = M^4 \left. \frac{d^2 V}{d \varphi^2}
	\right|_{\varphi_0} \sim \frac{M^4}{M_P^2}
\end{equation}
which must be positive at $\varphi_0 = 2 M^2 b_0 = 2 M_P$ to have a
linearly stable $b_0$ solution~\cite{hierarchy}.

The stability properties of $B_C(b)$ in Eq.~\eq{IIk-2} are the same as
of $B(b)$ in Eq.~\eq{II-2}: the respective solutions with $b(t_0) =
b_0$ and $\tilde{b}_0$ are linearly stable if the radion mass squared
is positive, 
since the radion mass squared including the
Casimir contribution
\begin{equation}
	m_{\varphi,C}^2 = m_\varphi^2 + \frac{5 k M^8}{M_P^6}
	\sim \frac{M^4}{M_P^2} \left( 1 + \frac{5 k M^4}{M_P^4} \right)
\end{equation}
is positive if $m_\varphi^2$ is, and are globally stable if the
respective potentials $B(b)$ and $B_C(b)$ are, for example, concave
upward (the simplest case), since
\begin{equation}
	B_C(b) = B(b) + \frac{k}{2 M^4 b^4} + {\rm const}
\end{equation}
is concave upward if $B$ is.

If the number of extra dimensions $D$ is allowed to be greater than
two, the Einstein equations~\eq{a} and~\eq{b} for an inflating 3-space
with static $b(t)$ change to
\begin{equation}
\label{a2}
	3 H_0^2 =
	-\frac{k}{M^{D+2} \tilde{b}_0^{D+4}} - \frac{D-2}{D+2} 
	\frac{V\left(\tilde{b}_0\right)}{M^{D-2} \tilde{b}_0^D} +
	\frac{V'\left(\tilde{b}_0\right)}
	{(D+2) M^{D-2} \tilde{b}_0^{D-1}}
\end{equation}
\begin{equation}
\label{b2}
	0 = 
	\frac{4 k}{D M^{D+2} \tilde{b}_0^{D+4}} + \frac{4}{D+2} 
	\frac{V\left(\tilde{b}_0\right)}{M^{D-2} \tilde{b}_0^D} -
	\frac{2}{D (D+2)}
	\frac{V'\left(\tilde{b}_0\right)}{M^{D-2} \tilde{b}_0^{D-1}}
\end{equation}
but the result for the present-day cosmological constant has the same
form
\begin{equation}
	\rho_\Lambda = \frac{k}{b_0^4} =
	\frac{k M^{4+\frac{8}{D}}}{M_P^\frac{8}{D}}
\end{equation}
where now $b_0$ satisfies $b_0^D M^{D+2} = M_P^2$.  Thus
$\rho_\Lambda$ has the right parametric dependence $M^8/M_P^4$ only
for $D = 2$.

\section{Conclusion}

The cosmological picture presented here joins smoothly onto the
primordial inflation and big-bang cosmological pictures: The quantum
birth of the universe begins with $a$ and $b \sim 1/M$.  Many ($\gg
70$) e-folds of primordial inflation occur as the inflaton $b(t)$
rolls down its potential to $\tilde{b}_0$.  $b(t)$ then undergoes
damped oscillations about $\tilde{b}_0$, heating the universe up to a
temperature $T$ above the temperature for big-bang nucleosynthesis
(BBN) and creating essentially all the matter and energy we see today.
(See Refs.~\cite{ADD-cosmo0} and~\cite{hyp-extra} for two differing
views on the maximum value of $T$, above which the evolution of the
universe in ADD-type theories cannot be described by the
radiation-dominated Friedmann-Robertson-Walker model.) At this point,
the universe evolves according to the standard big-bang picture,
expanding and cooling, with a fixed small cosmological constant
$\rho_\Lambda = k/b_0^4 \approx (2.3 \times 10^{-3}~{\rm eV})^4$.

This dark energy density is much less than the BBN energy density
$\sim (1~{\rm MeV})^4$ and plays a role in the evolution of the
universe only recently, long after the equality of energy density
$\sim (1~{\rm eV})^4$ in matter and radiation.  The radius $b(t)$ of
the compactified space has not changed since well before BBN.

Finally we note that if the stabilization potential $V(b)$ vanishes at
its global minimum, the resolution of the cosmic coincidences of
Ref.~\cite{cosmic-coincidence} is naturally realized in the Casimir
effect since parametrically $\rho_\Lambda \sim M^8/M_P^4$.

\end{document}